# Graphene Domain Signature of Raman Spectra of $sp^2$ Amorphous Carbons

*E.F. Sheka[1*], Ye. A. Golubev[2], N.A. Popova[1]*

[1] Peoples' Friendship University (RUDN University) of Russia, Miklukho-Maklay 6, 117198 Moscow, Russia

[2] Yushkin's Institute of Geology, Komi Science Center, Ural Branch of RAS, Pervomayskaya 54, 167982, Syktyvkar, Russia

*Corresponding author: E.F. Sheka
E-mail: *sheka@icp.ac.ru*

**ABSTRACT**. The paper presents a joint consideration of Raman spectra of $sp^2$ amorphous carbons alongside with the nature and type of their amorphicity. The latter was attributed to the enforced fragmentation. The fragments, presented with size-restricted graphene domains with heteroatom necklaces in the circumference, are the basic structural units (BSUs) of the solids, determining them as amorphics with molecular structure. The standard G-D-2D pattern of Raman spectra of polycyclic aromatic hydrocarbons, $sp^2$ amorphous carbons, graphene and/or graphite crystal is attributed to BSUs graphene domains. The molecular approximation allows connecting the G-D spectra image of one-phonon spectra with a considerable dispersion of the C=C bond lengths within graphene domains, governed by size, heteroatom necklace of BSUs as well as BSUs packing. The interpretation of 2D two-phonon spectra reveals a particular role of electrical anharmonicity in the spectra formation and attributes this effect to a high degree of the electron density delocalization in graphene domains.

## I. INTRODUCTION

Raman scattering has become an overwhelming method of testing graphene-based solid carbons. A comprehend review [1] could be recommended as a guide to the vast literature on the subject while providing in-depth view on the physics when dealing with Raman spectroscopy of graphene-based solids. The subject concerns two issues, each of which is highly peculiar and complicated. The first concerns the theoretical background of Raman spectroscopy in the case, while the second is associated with a broad meaning what is a 'graphene-based solid' under study. Thus, phonons of perfect as well as defect and disordered crystals and molecular vibrations are fundamentals of the first part forming the ground for two theoretical approaches, namely: solid-state and molecular ones (see Reviews [2, 3] and references therein). On the other hand, graphene-based solids, ranging from very well organized three coordinated $sp^2$ graphite, graphene, nanotubes, and nanoribbons, down to amorphous carbons, graphene quantum dots as well as various $sp^3$-$sp^2$ mixtures, and that is not all, present the rich content of the second part. Because of this, as shown [1], the modern Raman spectroscopy of graphene materials is a subtle art of a proper combination of the two components. The problem is particularly sharp when the solid is definitely nanostructured, which raises an evident question if either the phonon-based solid-state approach is applicable to the case or its molecular counterpart should be at play. The standard view of Raman spectra (RSs) of graphene-based species with main structural features presented by dominating D,

G, and 2D bands, have so far played a decisive role in the approach selecting. A deep similarity of the spectra for a large set of samples of defect and disordered crystals of graphite and graphene [4-9] has created a favorable basis for successful use and further improvement of the solid-state approach in its phonon-confinement format [10,11]. This format has legitimated the use of RSs parameters, such as position, bandwidth and intensity of the D, G, and 2D bands for the determination of the confinement parameters thus characterizing the carbon crystals nanostructuring (see a number of cases in [1] and references therein). The same standard appearance of the RSs of crystalline and non-crystalline graphene-based solids stimulated the spread of the phonon-confinement approach to amorphous carbons, graphene quantum dot, and other non-crystalline species (see [12-15], thus presenting the latter now as the main stream in the consideration of the RSs of all graphene-based carbons "proposed by international consensus" [1].

Nevertheless, recent detailed investigations have clearly revealed a molecular nature of $sp^2$ amorphous carbons (ACs) (*amorphics* for simplicity) [16, 17], again raising the question about the choice between solid-state and molecular approach. Meeting this request, in the current article we intend to inspect which new information about $sp^2$ ACs can be obtained from their spectra analysis based on the molecular spectroscopy background. To make the latter informative and convincing, we conducted a comparative study of RSs of a set of specially selected $sp^2$ amorphics, the structure and chemical composition of which were carefully investigated. This study allowed to make conclusion about the type of the amorphicity of $sp^2$ solid carbon, supporting its molecular character, and to reveal the fundamental difference between the molecular ACs and nanostructured graphite and graphene crystals subjected to size confinement.

The paper is composed as follows. Section 2 is devoted to the modern presentation of amorphous carbons. Specification of the ACs amorphicity is discussed in Section 3. General comments concerning vibrational spectroscopy of amorphous carbons as well as main concepts of molecular approach related to graphene molecules are considered in Section 4. Sections 5 and 6 present the interpretation of the obtained RSs in the framework of molecular approximation, concerning one- and two-phonon fractures, respectively. Conclusion summarizes main essentials received.

## II. TODAY'S PRESENTATION OF AMORPHOUS CARBON

Amorphous carbons are widely common and present a big allotropic class of bodies, both natural and synthetic. Natural amorphics are products of the activity of the Nature's laboratory during geological billion-million-year time. Exhausted geological examinations allowed suggesting a few classification schemes of carbon species, one of which, slightly changed with respect to original [18], is schematically shown in Fig. 1a. The scheme presents a continuous evolution of pristine carbonaceous masses into ordered crystalline graphite thus exhibiting the main stream of carbon life in the Nature. The evolution is presented as increasing carbonization rank of intermediate products. As seen in the figure, a general picture can be split into two gloves, the left of which starts with plants and sediments of different kinds and proceeds through sapropels to brown coals and later to convenient coals and anthracite. The endpoint on the way is graphite. As for the right glove, it covers carbonization of pristine gas and distillate oil and proceeds through petroleum and naphthoids to asphalts and then to kerites, anthraxolites, and shungites. As in the previous case, graphite is the endpoint. Certainly, the division is not exactly rigid, due to which a mixture of the two fluxes, particularly, at early stages of carbonization, actually occurs. This scheme is related to $sp^2$ amorphous carbon that actually dominates in the Nature. Natural $sp^3$ amorphous carbon is not so largely distributed, due to which diamond-like natural ACs have become top issues of the carbon mineralogy [19, 20] only recently. Natural amorphic sampling for the current study concerns $sp^2$ species including anthracite, anthraxolite and shungite carbon.

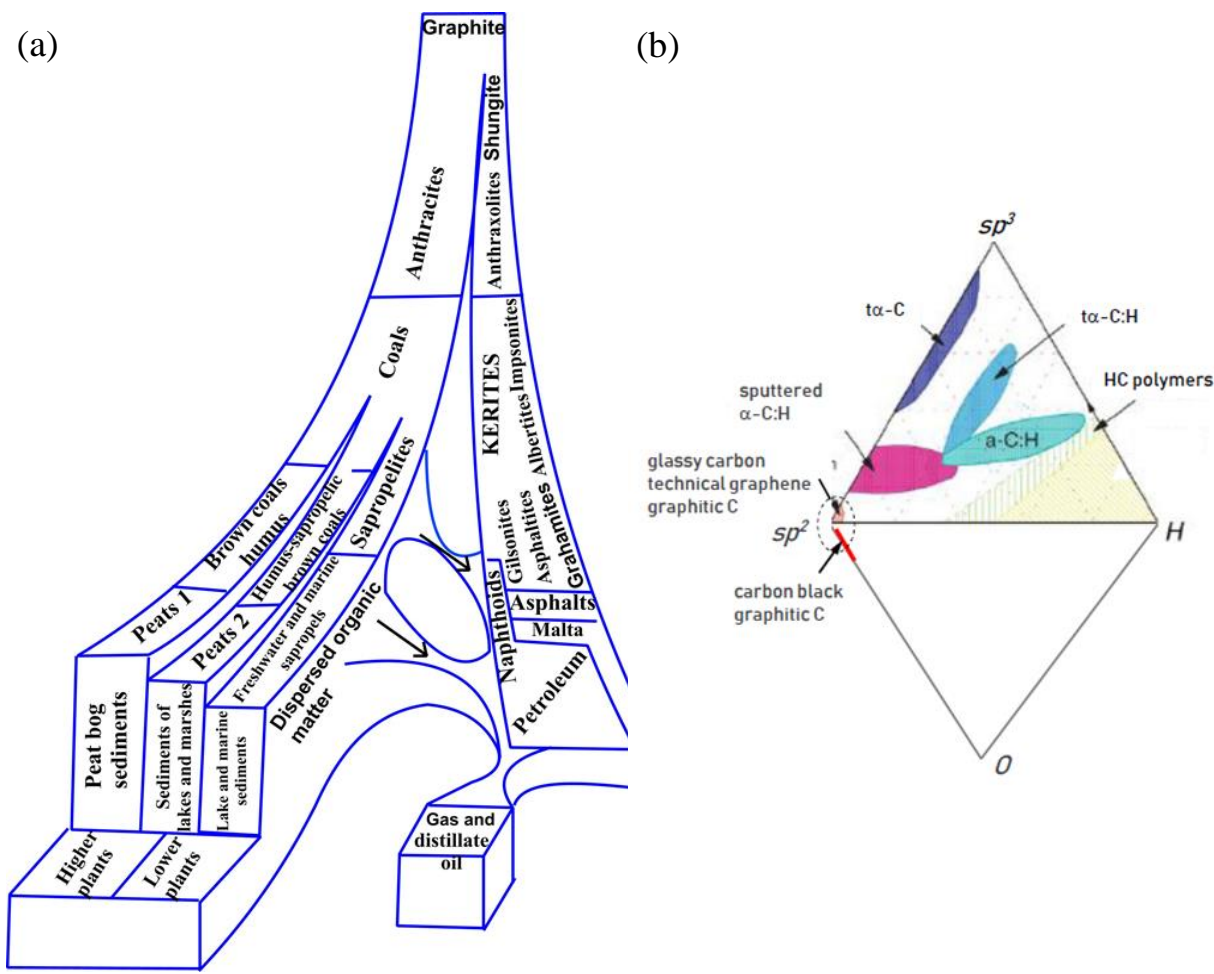

**Figure 1.** a. Path of carbon life in the Nature: Amorphous carbons based on the Uspenskiy's classification [18]. b. Rhombic phase diagram of synthetic amorphous carbon-hydrogen-oxygen system.

The family of synthetic amorphous carbons is quite large, covering species different by not only the carbonization rank, but also a mixture of $sp^2$ and $sp^3$ components. Analogously to natural, synthetic amorphics were classified as well [21, 22] and the relevant classification scheme is presented in Fig. 1b. Previously ternary, in the current study, it is completed to a rhombic one to take into account oxygen as one other important ingredient, which is caused by large development of modern techniques of the ACs production. Comparing schemes presented in Figs. 1a and b, it becomes evident that those are related to fully different communities of substances. If natural amorphics belong to $sp^2$ carbon family and are rank-characterized with respect to the stage of their metamorphism and carbonization, synthetic amorphics are mainly characterized by $sp^3$-configured solid carbon. $sp^2$ Group of the synthetic species takes only small place, marked by oval, in the total family and their carbonization is the highest. Because of a small amount of $sp^3$ solid amorphous carbon in nature, special technologies to produce tα-C, tα-CH (t means tetragonal), and sputtered $sp^2$ & $sp^3$ mixed α-C:H products were developed. It is necessary to complement this part of carbon solids by graphene oxide which, in ideal case, corresponds to $sp^3$ configured carbon only and to about 70 wt % of oxygen and which cannot be presented on the considered plane classification scheme. As for $sp^2$ synthetic amorphics, for a long time they were presented by multi-tonnage industrial production of glassy (covering graphitic, black, activated and other highly carbonized) carbon [23]. However, the graphene era called to life a new high-tech material – technical graphenes [24] which are the final product of either oxidation-reduction [25] or oxidation-thermally-shocked exfoliation [26] of nanosize graphite. Two new members of this community are on the way – that are laser-induced graphene (LIG) manufactured by multiple lasing on cloth, paper, and food [27] and extreme quality flash graphene (FG) [28], closest in ordering to graphite.

In the current study, our attention will be concentrated on *sp²* ACs only. Recent comparative studies [16, 17] together with a large pool of individual data have shown that these amorphics form a particular class of solids. As turned out, the general architecture of both natural and synthetic species is common and can be characterized as multilevel fractal one [29, 30], albite differing in details at each level. The amorphics's structure of the first level is well similar in all the cases and is presented by *basic structure units* (BSUs). The higher-level structure depends on the BSUs size. Thus, in the case of small-size natural amorphics, nanosize-thick stacks of nanosize BSUs present the second-level structure. These stacks form globules - a structure of the third level characterized with pores of the first nm. Further aggregation of globules leads to the formation of micro-nanosize agglomerates with pores of tens nm. Figure 2 presents schematically the evolution of such amorphic structure from a single BSU to macroscopic powder. Synthetic amorphics are characterized by a large dispersion of BSUs size from units to tens and/over first hundreds of nanometers. At the low-limit end of the dispersion, the amorphic structure is similar to that of natural species described above. At the high-limit end, the BSUs size does not prevent from BSUs packing in nanosize-thick stacks while the latter laterally extended are further packed in a paper-like structure.

According to studies [16, 17, 24], BSUs of both natural and synthetic *sp²* ACs present graphene molecules, which are elements of a honeycomb, or graphene domain, structure. Obeying the general laws of chemistry of nanosize objects [31], the domains edge atoms are terminated by heteroatoms and/or atomic groups, including hydrogen, oxygen, nitrogen, sulfur, halogens mainly. The molecules can be described by statistically averaged chemical formula (such as $C_{66}O_4H_6$ (or $C_6O_{0.36}H_{0.55}$ per one benzenoid unit) in the case of shungite carbon, a model BSU of which is shown in Fig. 2) that corresponds to the chemical content of the sample obtained experimentally. Size, shape, and 'chemical necklace' of BSUs, the latter includes terminating atoms and atomic groups at particular disposition in the molecule circumference, greatly vary, due to which each AC, classified usually by origin, covers a large class of specially framed graphene molecules. Nevertheless, despite the complexity of the overall fractal structure of *sp²* ACs, precisely BSUs determine the solids properties and stipulate molecular-structural approach for their description. Firstly successfully applied to vibrational spectra of the solids provided by INS and DRIFT spectroscopies [17, 32], in the current paper the approach is spread over the RSs. To this end, specific *sp²* ACs of the highest-rank carbonization were selected [16, 17]. This set involved samples of shungite carbon (ShC), anthraxolite (AnthX), and anthracite (AnthC), one chemically (Ak-rGO) and one thermal-shock (TE-rGO) exfoliated technical graphenes as well as two industrially produced Sigma-Aldrich carbon blacks 699632 (CB632) and 699624 (CB624), (see the detailed description of samples in [16, 17]). The set is complemented with two graphites of the best quality from Botogol'sk deposit [33] characterized by mono- (mncr) and micronanocrystalline (μncr) structure. The structural and chemical data of the samples are summarized in Tables 1 and 2. The data obtained earlier are supplemented in this study by the results of the X-ray diffraction and EDS measurements for the two graphites.

### III. THE AMORPHICITY OF *sp²* SOLID CARBON

Despite *sp²* ACs have been the object of study and practical use for hundreds of years, until now they have not been considered from the general concept of the solid-state physics. The amorphicity of solids was widely studied and the main concepts are accumulated in monograph [34]. The first conceptual issue concerns a considerable degree of amorphous solid ordering that is subdivided into short-range (local) and medium range ones, the boundary between which passes around a few nm. The second issue is related to the direct interconnection of the solids properties and their local ordering due to which establishing a local structure has always been the main goal of studying. Historically, the most attention has been given to monoatomic Si and Ge, which, it would seem, is quite conducive to our study, because a similar behavior could be expected to carbon as well. As was found, tetrahedrally bonded atoms form the short-range order of these solids. In the case of carbon, as seen in Fig. 1b, similar amorphous compositions of carbon are concentrated near $sp^3$ corner only and are related to t$\alpha$-C phase. The remaining part of the C-diagram is connected with the presence of $sp^2$-configured amorphics that is absent in the case of Si and Ge. We believe that this feature is on line with other ones related to the tetrel family, such as the absence of both aromatic families of Si and Ge chemical compounds and freestanding honeycomb monolayer silicene and germanene similar to graphene. The feature is caused by an extreme radicalization of the elongated double covalent bonds X=X (X= Si, Ge) [35, 36]. Accordingly, from the standpoint of the general concept of amorphicity, monoatomic solid carbon has the unique ability to form amorphous (as well as crystalline) states of two types, characterized by fundamentally different short-range orders presented by either tetrahedral groups of bonded atoms or a honeycomb network of benzenoid units, thus differentiating $sp^3$ and $sp^2$ amorphous carbon.

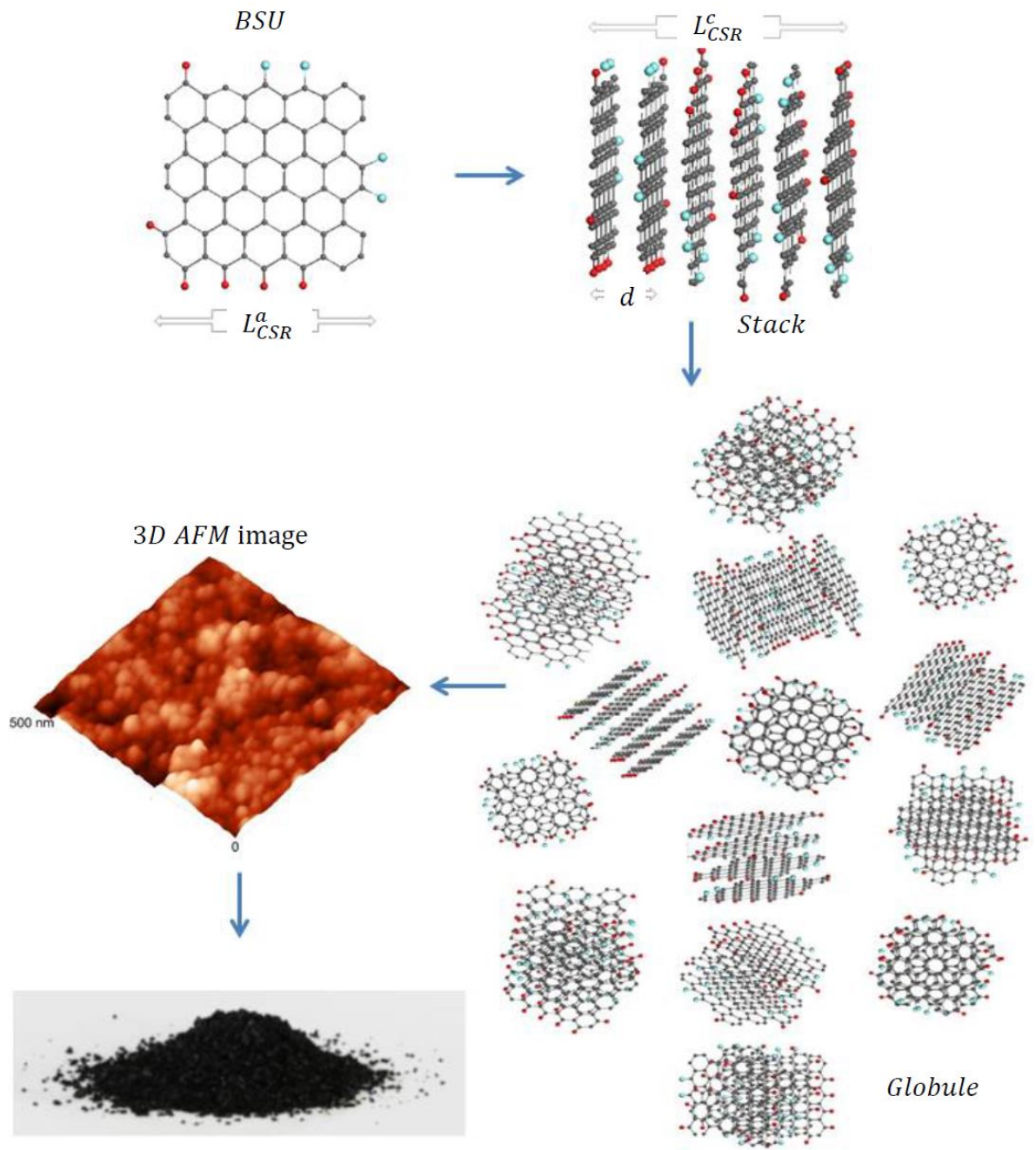

**Figure 2.** Schematic transformation from a single BSU molecule (one of the BSU models $C_{66}O_4H_6$ for shungite carbon [17]) to powdered solid amorphous carbon via BSU stacks and globule(s). The stacks consist of a number of BSU layers from 4 to 7, differently oriented to each other. Planar view on a model globule composed of different stacks, with total linear dimensions of ~ 6 nm. Dark gray, red and blue balls depict carbon, hydrogen, and oxygen atoms, respectively. 3D AFM image of globular structure of shungite carbon powder (NTEGRA Prima, NT-MDT) with 25 nm at the maximum of the nanoparticle size distribution.

**Table 1.** Structural parameters of amorphous carbons [1]

| Samples | Interlayer distance, $d$ (Å) | $L_{CSR}^c$, nm | Number of BSU layers | $L_{CSR}^a$, nm | Ref |
|---|---|---|---|---|---|
| mncr Gr[2] | 3.35(X) | 105 | 313 | 550 | this work |
| μncr Gr[2] | 3.35(X) | 49 | 146 | 184 | this work |
| ShC | 3.47(N); 3.48(X) | 2,5(N); 2.0(X) | 7(N); 5-6(X) | 2.1(X) | [16] |
| AnthX | 3.47(N); 3.47(X) | 2.5(N); 1.9(X) | 7(N); 5-6(X) | 1.6(X) | [16] |
| AnthC (Donetsk) | 3.50(X) | 2.2(X) | 5-6(X) | 2.1(X) | [16] |
| CB632 | 3.57(N); 3.58(X) | 2.2(N); 1.6(X) | 6(N); 4-5(X) | 1.4(X) | [16] |
| CB624[3] | 3.40(N); 3.45(X) | 7.8(N); 6(X) | 23(N); 17(X) | 14.6 | [16] |
| Ak-rGO | 3.50(N) | 2.4 (N) | 7(N) | >20(N)[4] | [37] |
| TE-rGO | 3.36(N) | 2.9 (N) | 8(N) | >20(N)[4] | [38] |

[1] Notations (N) and (X) indicate data obtained by neutron and X-ray diffraction, respectively.
[2] The data are obtained by the treatment of (002) and (110) reflexes using Scherrer's relation $L_{CSR} = K \cdot \lambda / B \cdot \cos\Theta$. Here $\lambda$ is the X-ray radiation wavelength (CuK$_\alpha$) 0.154 nm, $\Theta$ is the position of the (110) ($L_{CSR}^a$) and (002) ($L_{CSR}^c$) peaks, $B$ is the half-height width of the peak in 2$\Theta$ (rad) units, and constant $K$ constitutes 0.9 and 1.84 for reflexes (002) and (110), respectively, in the approximation of a disk-shaped particles.
[3] X-ray data were corrected in the current study.
[4] $L_{CSR}^a$= 20 nm means the low limit of the value accessible for the measurements performed.

As mentioned earlier, the short-range order of $sp^2$ ACs is provided with framed graphene molecules. The framing plays a decisive role, ensuring the formation and stability of short-range order, on the one hand, and preventing crystallization, on the other [29, 31], thus allowing to attribute the origination of $sp^2$ ACs to the *reaction amorphization* [34]. The medium-range order of the amorphics reliably follows from the fractal porous structure evidently observed experimentally [39] and is presented by nanosize BSU stacks, composed in either globules (natural amorphics [29]) or paper-like sets of stacks (technical graphenes [37, 38]). A model structure of the globule, shown in Fig. 2, corresponds to shungite carbon. Molecule $C_{66}O_4H_6$ constitutes one of possible models of the ShC BSUs based on a graphene domain $C_{66}$, related to chemical data of the species listed in Table 2 [17]. Composed into four-, five-, and six-layer stacks, the molecules create a visible picture of the medium-range order of the species. Medium-range-order globules of anthraxolite and anthracite can be similarly visualized using the relevant BSU models suggested in [17]. Apparently, similar medium-range configuration can be suggested for industrial carbon black CB632. Bigger lateral dimensions of BSUs of technical graphenes Ak-rGO and TE-rGO as well as carbon black CB624 make to think about laterally extended sets of several-layer stacks based on molecular BSUs similar to those presented in [17].

**Table 2**. Chemical content of amorphous carbons

| Samples | Elemental analysis, wt% | | | | | Ref. | XPS analysis, at% | | | Ref. |
|---|---|---|---|---|---|---|---|---|---|---|
| | C | H | N | O | S | | C | O | Minor impurities | |
| mncr Gr[1] | 99.0 | - | - | 1.0 | - | this work | | | | |
| μncr Gr[1] | 98.9 | - | - | 1.1 | - | this work | | | | |
| ShC | 94.44 | 0.63 | 0.88 | 4.28 | 1.11 | [26] | 92.05 | 6.73 | **S** - 0.92; **Si** – 0.20; **N**-0.10 | [16] |
| AnthX | 94.01 | 1.11 | 0.86 | 2.66 | 1.36 | [26] | 92.83 | 6.00 | **S** - 0.85; **Si** – 0.25; **N**-0.07 | [16] |
| AnthC | 90.53 | 1.43 | 0.74 | 6.44 | 0.89 | [27] | 92.94 | 6.61 | **Cl** - 0.11 - **S**: 0.34 | [16] |
| TE-rGO | 84.51 | 1.0 | 0.01 | 13.5 | 1.0 | [27] | 86.77 | 10.91 | **F** - 077; **S** - 0.86; **Si** -0.70 | [17] |
| Ак-rGO | 89.67 | 0.96 | 0.01 | 8.98 | 0.39 | [27] | 94.57 | 5.28 | **S** - 0.16 | [17] |
| CB624 | 99.67 | 0.18 | 0 | 0.15 | - | [26] | 95.01 | 4.52 | **Si** – 0.46 | [16] |
| CB632 | 97.94 | 0.32 | 0.04 | 1.66 | 0.68 | [26] | 93.32 | 6.02 | **Si** – 0.66 | [16] |

[1] EDS measurements

Continuing this brief discussion of the general view of the $sp^2$ carbon amorphicity, we should dwell on one more distinctive feature. Evidently, the molecular nature of BSUs makes it possible to reliably attribute $sp^2$ ACs to amorphics with a molecular structure. However, in contrast to typical representatives of this class of solids, which are based on molecules of a stable standard structure [40, 41], BSUs of $sp^2$ ACs are not standard and the data listed in Tables 1 and 2 represent only statistically averaged quantities. For example, the positions of hydrogen and oxygen atoms in the BUS circumference, shown in Fig. 2, can markedly vary. Besides, the transition from amorphous structure to ordered one occurs in different ways as well. In ordinary amorphics, this transition proceeds as a gradual ordering of the positions of the constituent molecules with respect to each other via a sequence of mesomorphic transformation thus aligning translational periodicity (see details in [41]). In contrast, the crystallization, better to say, graphitization of $sp^2$ ACs occurs due to increasing the BSU size, as is shown schematically in Fig. 3. To trace processing, we have fixed the atomic C:O:H relative content by the data related to shungite carbon and have simulated the intermediate transition by growing the graphene domain. Certainly, the size-increasing of individual BSUs is accompanied with bridging of both BSUs themselves and stacks of them. For many years, this bridging was considered as a main process of sequential ordering of the primary block-mosaic structure of natural ACs [42-45]. Obviously, the priority of one of the two processes depends on the environment. Nevertheless, not depending on whichever process dominates in practice, we should come to a conclusion that $sp^2$ amorphous solid state can not be attributed to any of the known types of disorder characteristic for monoatomic solids [34]. What is observed empirically in the case of ACs, in terms of solid-state physics can be attributed to *enforced fragmentation* of graphite and/or graphene crystals. Obviously, the fragmented solid can be obtained from both the top and bottom. In the first case, it concerns the disintegration of pristine graphite and/or graphene crystal while in the second case it concerns stopping the graphitization of pristine graphene lamellas.

There might be various reasons for fragmentation, including mechanical impact, chemical reaction, temperature shock, exposure to hard radiation, etc., which can be easily traced by the history of the production of the ACs samples. The fragments ensure the short-range order of the solids and are characterized by large variety with respect to not only different classes of ACs provided by different history and/or technology of their production, but the same class as well. The variety concerns the BSUs size, shape, variation of chemical content, and, which is the most important, the distribution of heteroatoms in the BSU framing area ('chemical necklace') at fixed atomic percentage in average. Thus, models presented in Fig. 3, are only 'one snapshot' of communities related to possible permutations of hydrogen and oxygen atoms in the framing area that have no number.

Another distinctive feature of $sp^2$ ACs lies in the radical nature of their BSU fragments. As shown [16, 17], BSUs of ~2 nm in size are molecular radicals whose chemical activity is concentrated at non-terminated edge carbon atoms (these atoms are clearly seen on the top of Fig. 3) [17]. However, as can be seen from the figure, increasing the molecule size leads to decreasing edge atoms number due to which, at a fixed atomic composition C: O: H, the number of non-terminated atoms and the degree of the associated radicalization decrease. Thus, it is possible to think that weight content of oxygen in graphites in Table 2 might correspond to a complete termination of the relevant BSUs. Simultaneously, the presence of oxygen in the best graphites itself (the finding should be typical for all graphites [16]) evidences the limited size of BSUs as well as the unavoidable termination of the latter in the circumference. As for the weakening of the radicalization of BSU as its size grows, the issue, in particular, explains the well-known empirical fact that enforced additional fragmentation leading to nanostructuring of initial $sp^2$ ACs significantly enhances the yield of the reaction when these amorphics are used as carbocatalysts [46, 47].

Raman spectroscopy of $sp^2$ ACs, as well as of other amorphous solids, is generally aimed at determining their short-range structure [34, 48]. However, the information obtained depends on theoretical motives laying the foundation of the spectra analysis. Thus, the solid-state approach sees the studied amorphics as carbon honeycomb fragments, which are presented by either flat or slightly curved [49] graphene sheets with standard C=C interatomic space of 1.42 Å in size [1]. In contrast, molecular approach suggests the consideration of these solids as honeycomb C=C covalent-bond compositions [50] with not fixed C=C bond lengths thus exhibiting their sensitivity to both environment and other conditions of the body's production and storage. Presented below is aimed at revealing how these expectations are met.

## IV. GENERAL CONCEPTS OF MOLECULAR APPROACH RELATED TO GRAPHENE MOLECULES

For vibrational spectroscopy to become a valuable short-range structural probe for $sp^2$ ACs, BSUs must be capable of being vibrationally excited independently from the surrounding amorphous matrix [34]. In the case of the studied bodies, the requirement is fully met due to the presence of the BSU framing areas, on the one hand, and the evident vibrational decoupling of the molecules weakly interacting with each other within the stacks, on the other. The next issue concerns the spectra internal content. In contrast to IR-inactivity of molecular vibrational modes, associated with covalent homopolar bonds, caused by nil static moment [51], the electronicpolarizability is quite favorable for these bonds. Respectively, IR photoabsorption and Raman scattering relate to different components of the ACs BSUs - the first reflects the structure and chemical composition of the BSU heteroatom necklaces, and this has been recently successfully demonstrated by studying DRIFT spectra of the amorphics under consideration [17], while Ramat scattering draws the signature of the BSU graphene domains.

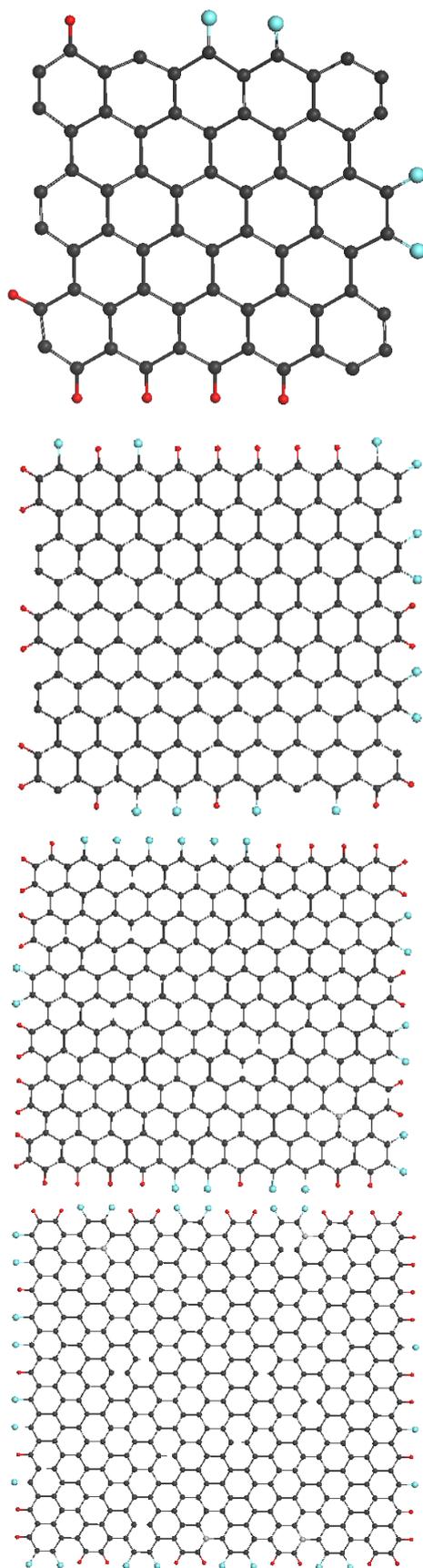

**Figure 3.** BSU models for sequential graphitization of amorphous carbon like shungite carbon. Graphene domains top down: (5,5)NGr, (9,9)NGr, (11,11)NGr, (15,12)NGr of lateral dimensions 1.2 nm, 2.2 nm, 2.7 nm, and 3.3 nm, respectively. Dark gray, red and blue balls depict carbon, hydrogen, and oxygen atoms. The chemical composition of the species is described by an average per one benzenoid formula $C_6O_{0.36}H_{0.55}$ in accordance with Table 2.

The molecular approach usually starts off a general consideration of the vibration spectrum of an object. Evidently, this spectrum of any graphene molecule is multitudinous and multimode due to which a certain simplification is needed to make it discussable. In the previous study [17] we suggested the spectrum of benzene molecule as the basis for simplified consideration of the vibrational spectra of $sp^2$ ACs BSUs. The list of benzene vibrational modes and their assignment are given in Table 3. According to symmetry rules, modes 1-10 are active in Raman scattering while modes 11-20 – in IR absorption. Evidently, any lowering of the molecule symmetry violates the double degeneracy and mixes the modes. Anyway, even with these limitations both Raman and IR spectra of benzene molecule should be quite rich. However, in practice both spectra of gaseous benzene are very simple and consist of small number of selected modes. Among the latter only $v_1$ and $v_7$ modes form the pattern of the observed RS. When the number of benzene rings increases, considerable growth of the vibrational modes occurs and the spectra patterns remarkably change. As for linear chains of benzene rings, the governing role in the spectra, starting from naphthalene, goes to stretching mode $v_8$ [52]. The corresponding band is surrounded by satellites due to the violation of the $D_{6h}$ symmetry of benzene and the removal of the degeneracy of the initial mode, first weak in naphthalene and anthracene, and then comparable in intensity in tetracene and pentacene [54]. At the same time, a scattering of noticeable intensity in the region of modes $v_8$ and $v_{19}$ arises in the spectrum of the last two molecules. Therefore, the transformation of RS of benzene when going to pentacene consists in enhancing the role of C=C stretching vibrations originated mainly from $e_{2g}$ $v_8$ and $e_{1u}$ $v_{19}$ benzene modes. The shape of the spectrum in this region is still rather complex.

Completely different feature is observed when going from polyacenes to polyaromatic hydrocarbons (PAHs) with two-dimensional (2D) $\pi$ conjugated planar structure. The latter represent spatially extended compositions of benzenoid rings forming restricted graphene domains framed by hydrogens. The first objects were PAHs $C_{78}H_{32}$ [55] and $C_{96}H_{30}$ [56] synthesized by Prof. Müllen's team. Raman spectra of both molecules [57] drastically differ from the spectra of polyacenes, taking form of a characteristic D-G-2D three bands pattern, which is pretty similar to the spectra of $sp^2$ ACs. Later on the PAH set was enlarged including molecules of different shape, symmetry, and carbon content from $C_{24}H_{12}$ to $C_{114}H_{30}$ [58, 59], whose RSs are of the same shape. Certainly, the characterization concerns the general pattern of the spectra, while $I_D/I_G$, and $\Delta\omega_D$ parameters were quite individual. Nevertheless, it was experimentally shown that just the presence of graphene domains in the 2D planar structure of PAHs leads to the characteristic D-G-2D shape of the RSs, which does not depend on the size and symmetry of the PAH molecules. The band triplet covers D-G one-phonon and 2D two-phonon parts of the spectrum.

A profound theoretical analysis of one-phonon RSs performed by the Italian spectroscopists [57-67] allowed both to reveal the reasons for the discovered uniqueness of the PAHs RSs, and to establish their intimate connection with the spectra of graphite and/or graphene. This analysis was based on a thorough study of the dependence of the polarizability tensor of the molecules, which determines RS intensity, on the dynamic characteristics of the molecules under conditions of a multimode structure of the vibrational spectrum and various point symmetries. It was found that the main contribution to the intensity of the spectra is made by C=C stretchings, due to which the observed D-G-2D set of bands is characteristic for the network of C=C bonds mainly. As for the stretchings themselves, the modes, which determine G band, are originated from the $e_{2g}$ vibration of benzene, while the modes responsible for D band come from the $e_{1u}$ vibration of the molecule (see Table 3). The modes individuality is caused by peculiarities of their vibrational forms. As convincingly shown (see Fig. 4), the vibrations of G band correspond to simultaneous in-plane stretchings of all C=C bonds, while those related to D band concern both stretching and contraction of these bonds when carbon atoms move normally to them just imitating benzenoid ring breathing. The motion of carbon atoms has a collective character, for which planar packing of benzenoid units is obviously highly preferable. In contrast, as seen in Fig. 4a, the vibrational forms of both $e_{2g}$ and $e_{1u}$ modes of benzene are local and differ much from collective forms of PAH molecules.

**Table 3.** Vibrational modes of benzene molecule (adapted from [52] )

| # | Symmetry | Frequency | Description |
|---|---|---|---|
| 1 | $a_{1g}$ | 993 cm$^{-1}$ | Breathing |
| 2 | | 3073 | C—H stretching in-phase |
| 3 | $a_{2g}$ | 1350 | C—H in-plane bend. in-phase |
| 4 | $b_{2g}$ | 707 | C—C—C puckering |
| 5 | | 990 | C—H out-of-plane trigonal |
| 6 | $e_{2g}$ | 606 | C—C—C in-plane bending |
| 7 | | 3056 | C—H stretching |
| 8 | | 1599 | C—C stretching |
| 9 | | 1178 | C—H in-plane bending |
| 10 | $e_{1g}$ | 846 | C—H out-of-plane (C$_6$ libration) |
| 11 | $a_{2u}$ | 673 | C—H out-of-plane in-phase |
| 12 | $b_{1u}$ | 1010 | C—C—C trigonal bending |
| 13 | | 3057 | C—H trigonal stretching |
| 14 | $b_{2u}$ | 1309 | C—C stretching (Kekulé) |
| 15 | | 1146 | C—H in-plane trigonal bending |
| 16 | $e_{2u}$ | 404 | C—C—C out-of-plane bending |
| 17 | | 967 | C—H out-of-plane |
| 18 | $e_{1u}$ | 1037 | C—H in-plane bending |
| 19 | | 1482 | C—C stretching |
| 20 | | 3064 | C—H stretching |

Polarization of molecules is highly sensitive to the vibration form. Actually, the quantity is generally described as [52]

$$\alpha_t = \alpha + \sum_i \left(\frac{\partial \alpha}{\partial Q_i}\right)_0 Q_i + \frac{1}{2!}\sum_{ik} \left(\frac{\partial^2 \alpha}{\partial Q_i \partial Q_k}\right)_0 Q_i Q_k + \cdots \qquad (1)$$

where α is the polarizability in equilibrium position, $Q$'s are co-ordinates of individual normal vibrations, sets $\{Q_i\}$ and $\{Q_k\}$ present vibration forms of the $t^{th}$ vibration, and the subscripts $_0$ of the differentials refer to the equilibrium position. The third and subsequent members of the power series represent the electrical anharmonicity. The intensity of one- and two-phonon Raman scattering is governed by the second and third terms, respectively. As seen from the equation, the vibration forms are directly involved into the intensity determination, differently for the Raman scattering of the first and second order.

Detailed consideration of the PAHs polarizability performed in the extensive study [57-67] showed that parallel-to-bond vibration forms, attributed by the authors to the type Я [59]), promote a steady intense one-phonon Raman signal G in all the studied molecules. The feature does not depend on the molecules symmetry and shape and is caused by the $\frac{\partial \alpha}{\partial Q_i}$ derivatives, which all are positive. In contrast, normal-to-bond vibration forms of type A [59] promote in this case both positive and negative $\frac{\partial \alpha}{\partial Q_i}$ derivatives due to which the intensity of D band is tightly connected with the "quality" of C=C bonds which is reflected in the vibration forms. So, if the bonds are identical, the contribution of A modes into the one-phonon signal is nil. In the opposite case, the signal is not nil and is the bigger, the bigger the difference between the bonds [60]. This conclusion has been approved on a number of PAH molecules, whose RSs were obtained and analyzed [57-60, 65, 67] as well as by latest theoretical studies [50]. Coming back to benzene, it becomes evident that the lack of collective character of the vibration forms of $e_{2g}$ and $e_{1u}$ modes (see Fig. 4a) prevents from the modes appearance in the molecule RS. Moreover, due to identity of C=C bonds and symmetry rule, D band must not be observed at all while G band of small intensity might be seen that is the real case.

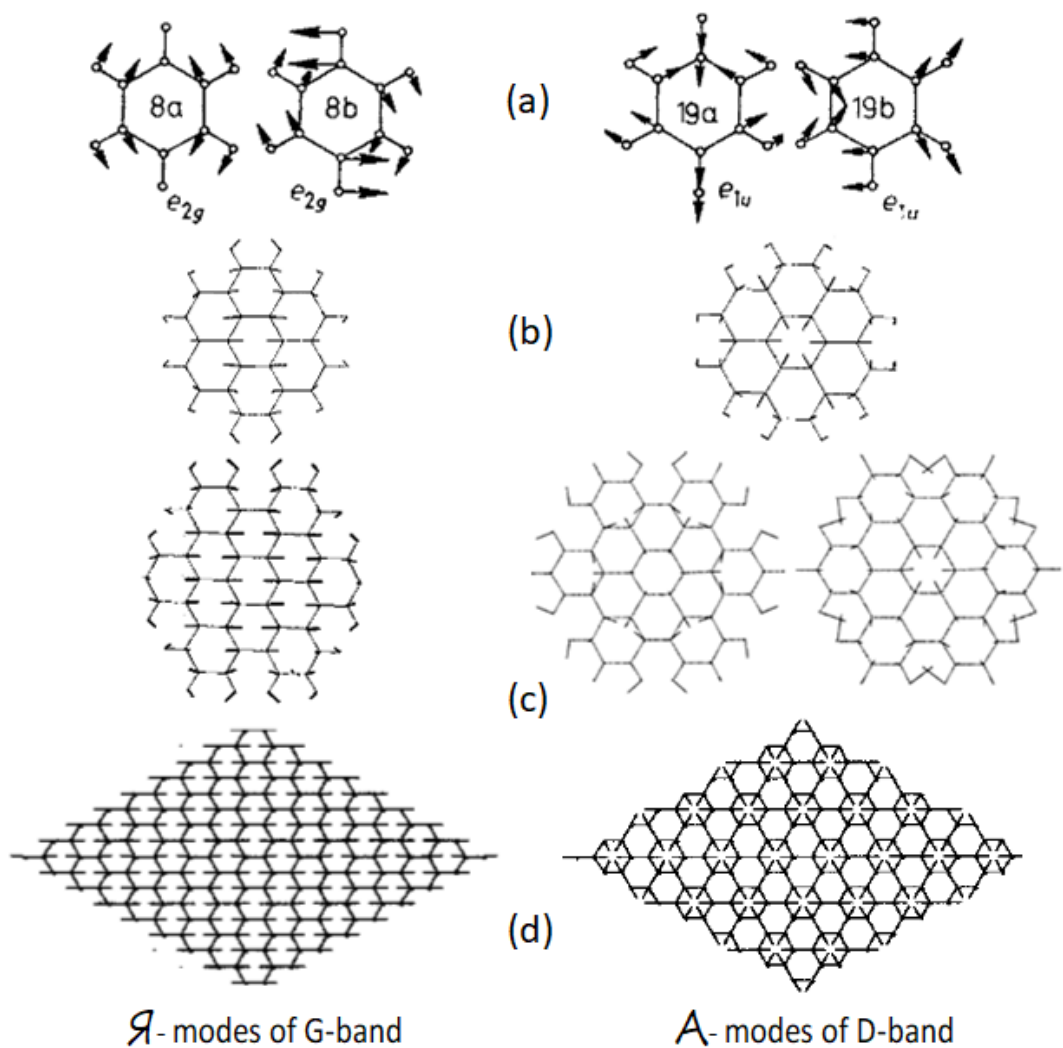

**Figure 5**. Vibrational forms of selected normal vibrations: (a) benzene; (b) coronene; (c) hexabenzocoronene; (d) a perfect 2D lattice of graphene. Adapted from [52] and [60].

From the very beginning of the study, Italian researchers have borne in mind RS of graphite and/or graphene crystals as a final point of the PHAs evolution by size. They extended their consideration over graphene crystal and have found that $e_{2g}$ mode at $\Gamma$ point of the first Brillouin zone is typical Я-mode while $A'_1$ mode at $K$ point clearly reveals A character (Fig. 5d). The findings undoubtedly evidence a peculiar collective character of the graphene phonons caused by benzenoid-hexagon structure. Therefore, until the honeycomb packing of benzenoid units is not broken, the graphene molecules and/or supramolecules are characterized by the D-G patterned Raman spectra. Only a complete destruction of the sheet leads to this feature loosing, which was really observed for graphene after its extremely strong bombardment by $Ar^+$ ions [68].

Figure 4, in general, is deeply consonant with Fig. 3 and can be considered as a RS image of sequential graphitization of amorphous carbon. Bearing this in mind, let us analyze RSs of the studied ACs in details.

## VI. ONE-PHONON RAMAN SPECTRA OF *SP²* AMORPHOUS CARBONS

Despite RSs of *sp²* ACs were recorded countlessly (suffice to say that Raman scattering is an indispensable participant in testing each natural and synthetic graphene material), a detailed investigation of a set of different-origin representatives of these materials under the same level of knowledge about their structure and chemical content has not yet been performed. In the current study, nine products of such kind described in Section II were examined. Raman spectroscopy was carried out with a LabRam HR800 instrument (Horiba, Jobin Yvon) at room temperature. The system was equipped with an Olympus BX41 optical microscope and a Si-based CCD detector (1024 × 256 pixels). A 50× objective (working distance ~3 mm, numerical aperture 0.75) was used. Spectra were recorded in the 100–4000 cm$^{-1}$ range using a spectrometer grating of 600 g/mm, with a confocal hole size of 300 μm and a slit of 100 μm. External laser exciting radiation in the region 270-520 nm was used. The obtained RSs reveal the expected dependence on the radiation frequency typical for nanostructured graphite and or graphene (see [1] and references therein) as well as for PAHs [50]. Figure 5 presents a collection of RSs of the studied samples excited by the radiation of Ar+ laser (514.5nm, 1.2 mW). This radiation is used in the majority of Raman scattering experiments, which makes it possible to compare the obtained spectra with the data available in the literature. Each spectrum in the figure is the result of three accumulations with a 10 s exposure. The spectra decomposition, if necessary, is provided with LabSpec 5.39 program using a pseudo-Voigt function (Gaussian-Lorentzian Sum) basis.

The spectra are grouped in two columns related to natural (left) and synthetic (right) samples. Looking at this collection, we would like to start with the first features that concern fine-structured spectra located in the first row in the figure. Among the latter, there are two lineaments, which require a particular attention. The former is related to the RSs of graphites while the latter – to the RS of amorphous carbon black CB624. As for graphites, a scrupulous analysis of the available RSs reveals that a single G-band-one-phonon spectrum is a rarity, once characteristic for "the best" graphites such as Madagascar flakes and Ticonderoga crystals [4], Ceylon graphites [69], Botogol'sk graphites [69, 70], and some others. In contrast, in the predominant majority of cases, researchers are dealing with highly structured graphites with a characteristic D-G doublet pattern of their RSs. Only high temperature and pressure could promote efficient graphitization in the Nature that is why large pieces of monocrystalline graphite are usually formed at great depth in the Earth core, as is the case of Botogol'sk deposit [33]. However, even under such conditions graphite rocks are not structurally homogeneous, consisting of microscopic monocrystalline (mncr) blocks surrounded with micronanocrystalline (μncr) graphite domains. Graphite spectra in Fig. 5a exhibit these two constituents. Expectedly, BSUs of the studied graphites are of submicron lateral size (see $L^a_{CSR}$ data in Table 1), once terminated mainly by oxygen, the weight content of which constitutes generally ~ 1 wt% (see Table 2). Similarity of the spectrum of amorphous CB624 and that of the μncr graphite convincingly evidences that not only in the depth of the Earth core, but also in industrial reactors producing carbon black a significant graphitization of amorphous carbon may occur. D-G doublet of narrowband structure in μncr graphite and CB624 amorphic, becomes the dominant feature of RSs of other studied ACs, both natural and engineered ones, but significantly broadened.

In the covalent-bond language of graphene molecules, the conversion of a single G band, corresponding to C=C stretching of an ideal graphite crystal, into a broadband D-G doublet of amorphous solids, while maintaining a honeycomb composition in the arrangement of carbon atoms, is caused by two things that result from the crystal fragmentation. The first concerns the genesis of non-zero dispersion of the C=C bonds length (*bond length dispersion* (BLD) below),

$\Delta\{l_{CC}\}$, in restricted graphene domains. Accordingly, a solitaire optical G phonon of both graphite and graphene crystals exists in a non-fragmented structure with a strictly fixed C=C bond value at 1.42 Å only. The second is a consequence of the graphene domain symmetry lowering just resolving not only $e_{2g}$, but $e_{1u}$ benzene-naming C=C stretchings discussed in the previous Section. Each of these groups covers a large number of C=C modes in polyatomic molecules like PAHs thus splitting the total BLD into two sets, $\Delta\{l_{CC}\}_{e_{2g}}$ and $\Delta\{l_{CC}\}_{e_{1u}}$, related to G and D components of the doublet spectra of PAHs, respectively. As was said earlier, the former modes are of Я type and are steadily active under Raman scattering, thus providing the appearance of G bands. The latter, revealed as D band only in the presence of C=C BLD, are of A type [3].

According to the vibrational dynamics of molecules, dispersion of any valence bond length $\Delta\{l_i\}$ is naturally transformed into that of force constants $\Delta\{f_i\}$ related to the relevant stretching. To evaluate the BLD value related to the C=C stretchings of the studied amorphics BSUs, we perform quantum chemical calculations of the ground state structure of a set of BSU models related to commensurate small-size ACs with $L_{CSR}^a$ of 1.4 - 2.1 nm (see Table 1) [17]. All the models, shown in Fig. 6, have the same graphene domain consisting of 66 carbon atoms, of which only one (III), two (IV), and four (II) atoms are substituted by oxygen. The structure set is complemented with the C=C bond length distribution related to each molecule. As seen in the figure, these distributions are quite similar for all the molecules albeit remarkably varying in response to varying compositions of the models 'chemical necklaces' thus evidencing a different assortment of both bonds and vibrational frequencies of each amorphic. As seen in the figure, the C=C bond lengths cover region from 1.5 Å to 1.3 Å in all the cases, thus determining the total dispersion of the bonds as $\Delta\{l_{CC}\} \geq 0.2$ Å. Without exact solution of the dynamical problem for the molecules we cannot distinguish Я and A modes as well as $\Delta\{l_{CC}\}_{e_{2g}}$ and $\Delta\{l_{CC}\}_{e_{1u}}$ BLDs separately and must limit ourselves by the discussion of the total value $\Delta\{l_{CC}\}$. Following Badger's rule [71], vibrational frequency and bond length of C=C stretching are interconnected so that $f_i = 3.0(l_i + 0.61)^{-3}$ [72] for constant and length units in $10^2$ Nm$^{-1}$ and $10^{-10}$ m, respectively. According to this relation, changing the length $l_{CC}$ from 1.5 Å to 1.3 Å corresponds to frequency growing from 1300 cm$^{-1}$ to 1700 cm$^{-1}$, which completely covers the frequency region of the D-G plottings for the studied solids.

The discussed above allows drawing a general conclusion that the D-G doublet structure of the RSs of *sp$^2$* ACs as well as their broadbandness are caused by the length dispersion of the C=C bonds that configure graphene domain structure of the relevant BSUs. There might be a few reasons for the dispersion. First, as said before, the fragmented character of the BSUs, once being graphene domains confined within a molecular size. Second, the BSU mandatory termination by heteroatoms, which clearly exacerbates the latter thus, making it dependent on a BSU size and the relevant 'chemical necklace'. Besides, it is known [73, 74] that multilayer packing of graphene sheets significantly influences RS of each of them as well, pointing to the additional redistribution of C=C bonds in the layers. Therefore, Raman scattering features of *sp$^2$* ACs concern two first levels of their multilevel structure, namely, BSUs and stacks of them. Apparently, there are other factors affecting RS, however, already these three factors are enough to imagine the complexity of the Raman scattering interpretation in this case. In what follows we will try to analyze the obtained RSs in terms of the empirical triad involving size and chemical necklace of BSUs as well as their packing.

Let us start from the RSs of natural amorphics in Figs. 5 b-d, BSUs of which are commensurate with PAH $C_{114}H_{30}$, RS of which was carefully analyzed [60]. As should be expected, the spectra of amorphics and PAH are well similar, albeit differing in details. The PAH spectrum consists of well isolated D and G bands with FWHM of 30 cm$^{-1}$ due to which the gap between the band is well pronounced. The D band is accompanied with four clearly distinguished low-intense satellite bands around the main peak. The satellite surrounding is characteristic for the D band in all the studied PAHs [51, 57, 67]. As for amorphics spectra, the FWHM of both main D and G peaks is doubled comparing with that of the PAH, while the satellite escort of the PAH D band is transformed into marked pedestal of different fine structure for amorphics significantly broadening the band in contrast to the G band.

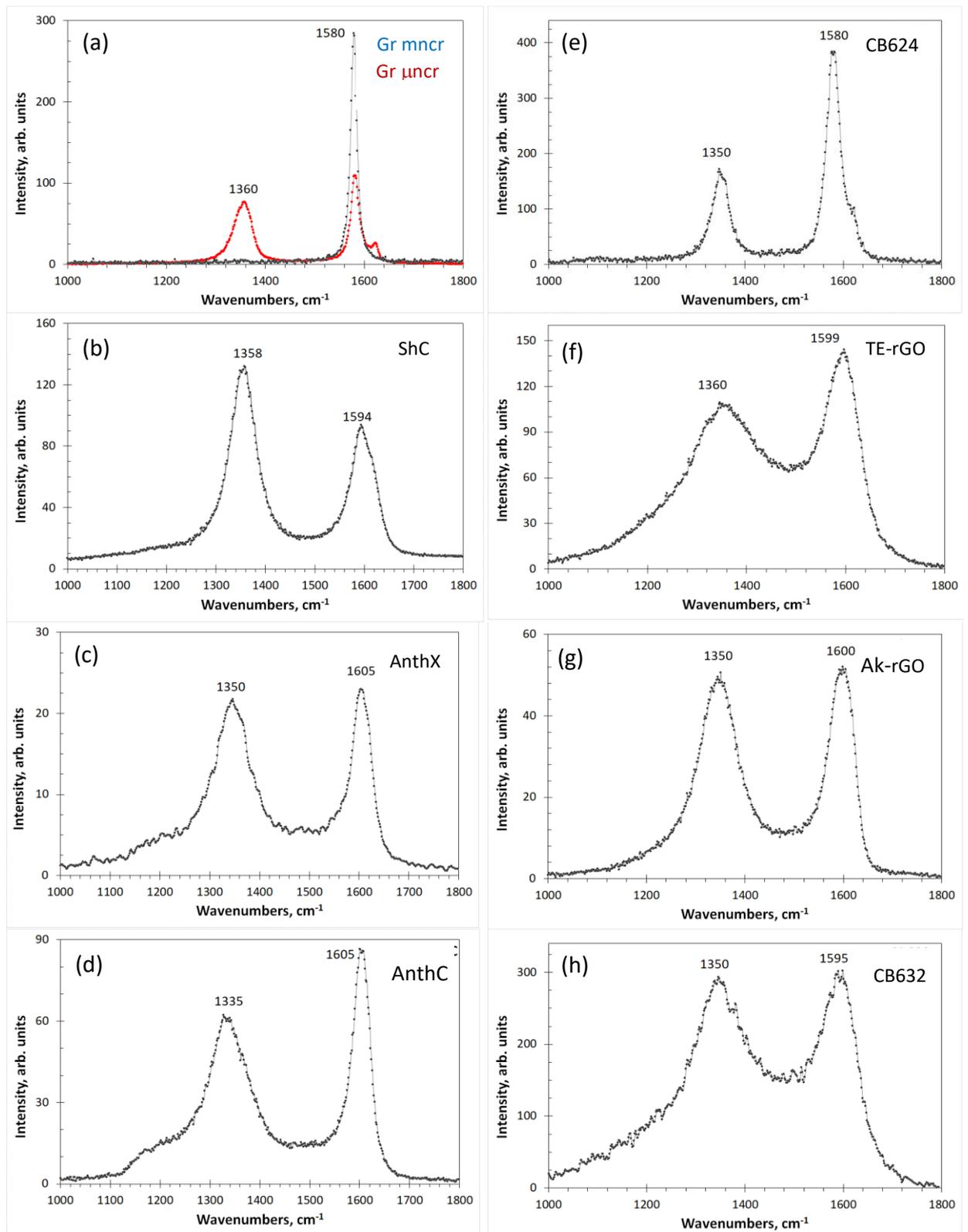

**Figure 5**. One-phonon Raman spectra of $sp^2$ amorphous carbons at room temperature: shungite carbon (ShC), anthraxolite (AnthX), anthracite (AnthC), technical graphene TE-rGO [38] and Ak-rGO [37], carbon blacks CB632 and CB624, as well as mono (mncr) - and micronanocstructured (μncr) graphites, respectively

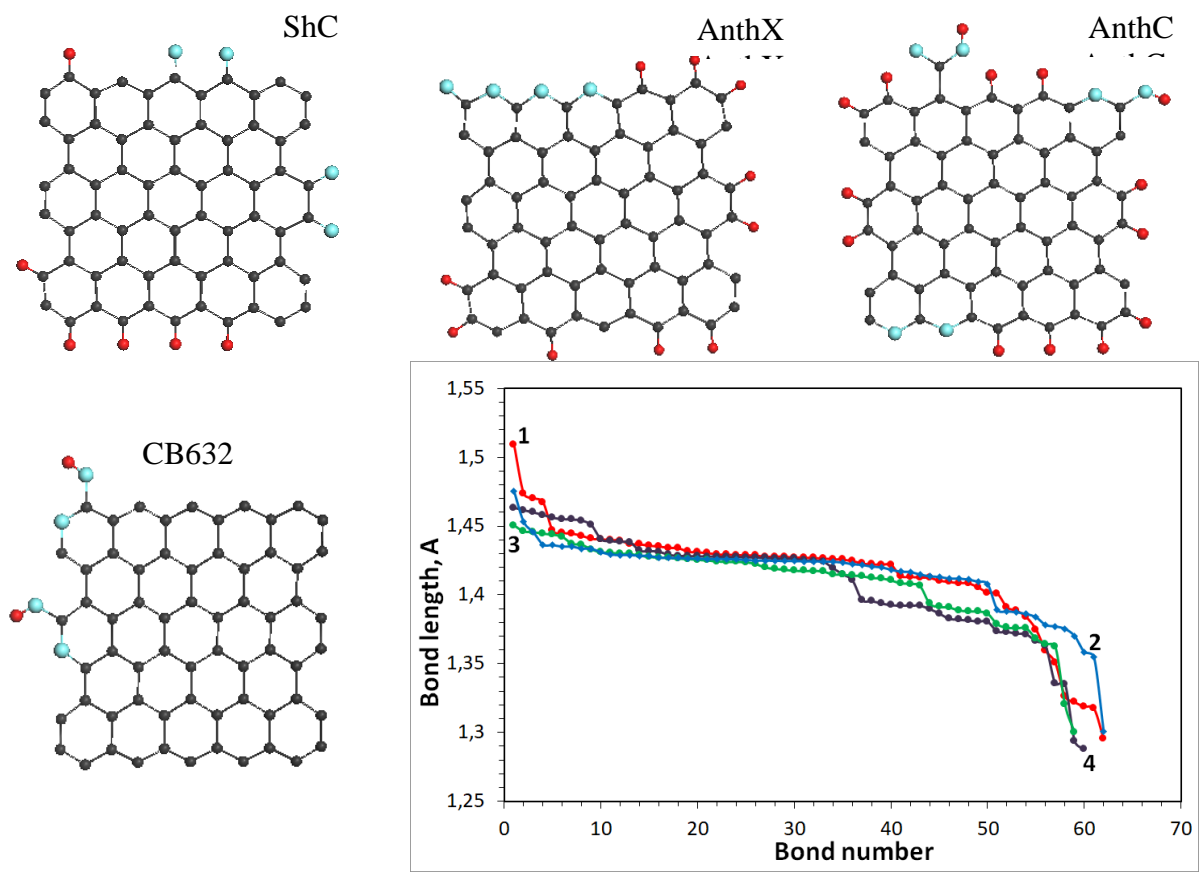

**Figure 6**. Equilibrium structure of BSU models related to shungite carbon (I), anthraxolite (II), anthracite (III), and carbon black CB632 (IV). Distribution of C=C bond lengths of the model graphene cores. UHF AM1 calculations.

In the light of the analytical triad described above, the spectrum of the PAH molecule contains information about the size of the graphene domain and the termination of the dangling bonds of its edge atoms with hydrogen. A detailed theoretical analysis of the spectra of similar PAHs was performed in terms of the totality of its C=C bonds in the presence of terminating hydrogen atoms [67], however, the contribution of the latter to the RS shape was not distinguished. Since, nevertheless, the spectrum of the $C_{114}H_{30}$ molecule consists of narrower bands than the spectra of the amorphics under consideration, it is quite reasonable to take it as the reference spectrum of a graphene domain of 1.5–2 nm in size. Respectively, the observed additional broadening of the amorphics RSs can be attributed to either heteroatom necklace of their BSUs or the BSUs packing.

Neutron and X-ray powder diffraction of shungite carbon and anthraxolite showed that the configurations of the solids stacks formed by the relevant BSUs were identical and the closest to the packing in graphite crystal (see [17] and references therein). Accordingly, these solids RSs broadening with respect to that of the PAH can be attributed to stronger effect of the BSU heteroatom necklace on the C=C BLD $\Delta\{l_{CC}\}$ in comparison with that of hydrogen atoms. Additional broadening can be caused by the variety of the necklace structure at fixed chemical content. Therefore, the difference of the ShC and AntX RSs in Figs. 4b and c can be attributed to the different disturbance of the BLD $\Delta\{l_{CC}\}$ by different heteroatom necklaces, which is supported with bond length distributions shown in Fig. 6. Similarly, the difference of the AnthC RS in Fig.

5d from the spectra of ShC and AntX can be attributed to the changes in the relevant necklace. Nevertheless, the structure study of the latter revealed much stronger deviation of BSUs packing from the graphite one [17] due to which the RS shape, particularly in the D band region, may additionally reflect the changing of BSUs packing in this case towards turbostratic one [75].

To the most extent, the influence of the turbostratic packing is seen when comparing the RS of carbon black CB632 in Fig. 5h with the spectra of natural amorphics discussed above. As seen in the figure, the CB632 spectrum drastically differs from those of natural species despite all the BSUs are commensurate and the influence of the BSU heteroatom necklace is also comparable (see Fig. 6). At the same time, structural data clearly evidence the turbostratic packing of the BSUs in this case (see [32] for details), which causes so strong broadening of the spectrum thus manifesting the effect of neighboring layer on the C=C BLD in each individual layer.

Basing on the conclusions, made when analyzing RSs of small-size natural amorphics and carbon black CB632, we can proceed with the interpretation of RSs of technical graphenes Ak-rGO and TE-rGO as well as carbon black CB624, related to large-size amorphics. As seen in Table 1, amorphous technical graphenes are characterized by large BSUs of submicron size, bigger than that one of carbon black CB624. Nevertheless, their RSs in Figs. 5f and g differ drastically from the spectrum of the latter in Fig. 5e clearly evidencing strong disordering and large BLD $\Delta\{l_{CC}\}$ in the relevant graphene domains. Important to note that the two spectra strongly differ in between as well. At the same time, the Ak-rGO spectrum is similar to that of AntX, while the same can be said about the spectra pair of TE-rGO and CB632. In both cases, the similarity is observed despite more than one order of magnitude difference in the BSUs size.

Discussing RSs of natural amorphics, we referred to the spectrum of the PAH $C_{114}H_{30}$ as the reference related to small-size graphene domains. In the case of large-size domains, evidently the spectrum of CB624 in Fig. 5e can play the role. The comparison of the Ak-rGO spectrum with the reference reveals doubling of FWHMs for both D and G bands at maintaining the general D-G appearance of the spectra as a whole. The feature is similar to that one resulted from the comparison of the PAH $C_{114}H_{30}$ and AnthX spectra. Since the additional broadening of the AnthX spectrum, which is similar to that of Ak-rGO, we attributed to the changing of the BLD $\Delta\{l_{CC}\}$ caused by complicated heteroatom necklace of the relevant BSUs, it is reasonable to replicate this conclusion with respect to the Ak-rGO–CB624 pair of spectra. The model necklace related to Ak-rGO amorphic, suggested on the basis of extended neutron scattering and DRIFT studies [17], is shown in Fig. 7a (left). As seen in the figure, it is quite cumbersome while causing the broadening comparable with that of AnthX. Additionally, despite the necklace is much complex with respect to that of AnthX, the D band in the RS of Ak-rGO is more symmetric without traces of the satellite surrounding structure. Similarly symmetric is the D band in the RS of CB624 in Fig. 5e, which apparently is resulted from the large size of the relevant BSUs. In contrast to Ak-rGO, RS of the TE-rGO is drastically different despite the commensurate BSU and similarity of the complex chemical necklace (see Fig. 7a (right)). A close resemblance of RSs of this amorphic and CB632, discussed earlier, leads to the conclusion that in both cases, a particular packing of the BSU layers is responsible for the spectra broadening. Actually, as in the case of CB632, neutron and X-ray powder diffraction reveals the turbostratic packing of paper-like sheets of TE-rGO [38], the consequences of which are clearly visible in Fig. 7b when comparing external view of the Ak-rGO and TE-rGO solids.

The analysis performed above convincingly shows that the C=C bond length distributions is responsible for the complicated broadband structure of the observed D-G spectra of $sp^2$ ACs. As seen in Fig. 6, the distribution is not homogeneous over the length scale and can be grouped. Evidently, grouping of C=C bonds in the studied amorphics causes a similar response of the C=C stretchings frequencies thus laying the foundation of 'multiband' origin of their broadband D-G RSs. To stress the attention on such grouping, Fig. 8a presents the bond length distributions, shown for exemplary models in Fig. 6, in different way. As seen in the figure, C=C bonds actually form distinguished groups. It is natural to expect that in the real RSs each of these groups should be associated with a relatively narrow band, because of which the observed spectrum represents a convolution of such bands.

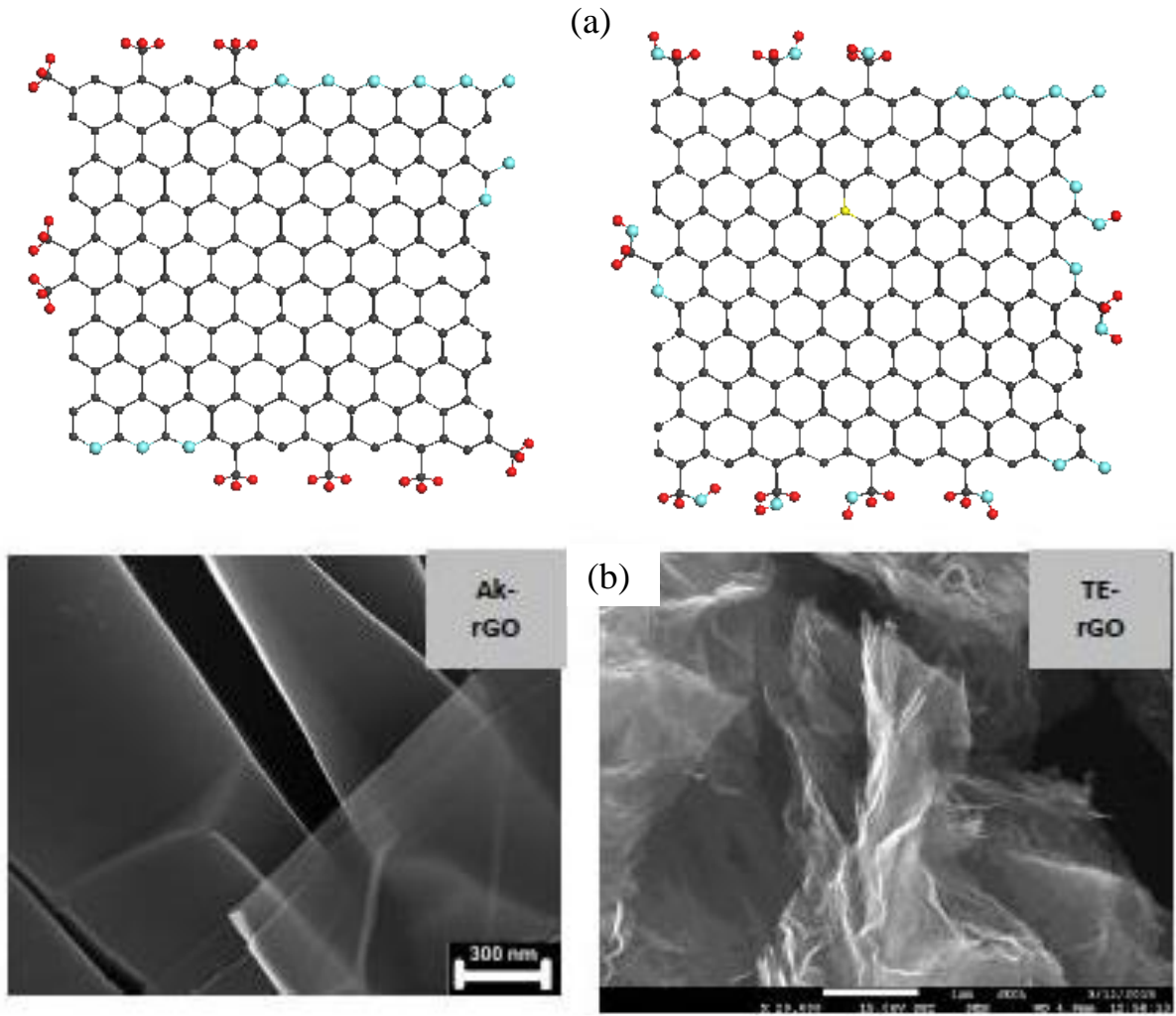

**Figure 7**. a. Molecular models of basic structure units of technical graphenes Ak-rGO (left) and TE-rGO (right). Light gray, red and black circles mark carbon, oxygen and hydrogen atoms. b. SEM images of technical graphene Ak-rGO and TE-rGO. Adabted from Ref. [17].

Intuitively, the multiband character of the D-G spectra was taken into account and adopted by spectroscopists from the first studies of Raman scattering by amorphous carbon and graphites (see, for example, [70, 76] and references therein). Standard programs of spectra deconvolution, such as LabSpec 5.39 in the current study, were widely used for experimental spectra decomposition. A typical example of such treatment is presented in Fig. 8b. The set of D1-D4 bands complemented with G band presents the usual basis for decomposing. Quite narrow spectral region and governing role of D1 and G bands provide a rather small dispersion in the maximum positions and FWHMs of D2-D4 bands thus giving a possibility to use the D1-D4 and G bands features when comparing RSs of different samples. Until now, it has been a comfortable formal language only facilitating the spectra description, while any changing of the D-G spectrum shape directly exhibits the reconstruction of the C=C bonds sets of the samples BSUs graphene domain.

Concluding the analysis of one-phonon D-G spectra, one should focus on the connection of widely used parameter $I_D/I_G$ with the real size of graphene domains. The advantage of the current study is the availability of independent data on the size of the studied BSUs obtained previously [16, 17] and presented in Table 1. Basing on these data, we come to unexpected results. So, for natural amorphics (spectra in Figs. 5b-d), the BSU sizes of which are approximately the same, the $I_D/I_G$ doubles. The parameter for CB632 (Fig. 5h) of practically the same lateral size is close to that for AnthX, but differs from both ShC and AnthC. The same can be said about the parameters for spectra CB624, TE-rGO, and Ak-rGO (Figs. 5e-g) with commensurate size parameters. Thus, the parameter $I_D/I_G$ can hardly be considered as an indicator of the graphene domains size in this case. The same conclusion was reached by the Italian researchers on the analysis of the D-G spectra of PAHs [59]. At the same time, on the experimental field of Raman spectroscopy of materials with elements of a graphene structure, there is a widespread belief in the unambiguous relationship of this parameter with the size of the graphene domain. At the time, this statement was confirmed by joint studies of the structure (X-ray diffraction) and Raman spectra with the determination of the ratio for nanostructured graphite and graphene crystals (see [6-9] and references therein). The first $I_D/I_G$ via $L_a$ relation proposed by Tuinstra and Koenig [4], gradually changed its appearance, supplemented by new parameters. The greatest recognition in practice has received the relation suggested by Consado et.al. [77]. Based on spatial confinement model applied to phonons in disordered graphene-based carbons [10, 11], this approach works quite satisfactorily in the cases of microstructured solids. Encouraged by this success, practitioners of Raman spectroscopy began to use these relationships in the case of nanostructured objects as well (see [76] and references therein), for most of which there are no independent structural data. Actually, only in this work such data are presented, which allows us to verify the validity of using the ratio to determine $L_a$. Table 4 shows the $L_a$ data obtained by processing the spectra shown in Fig. 5 using the relations $I_D/I_G$ via $L_a$ from [4] and [77].

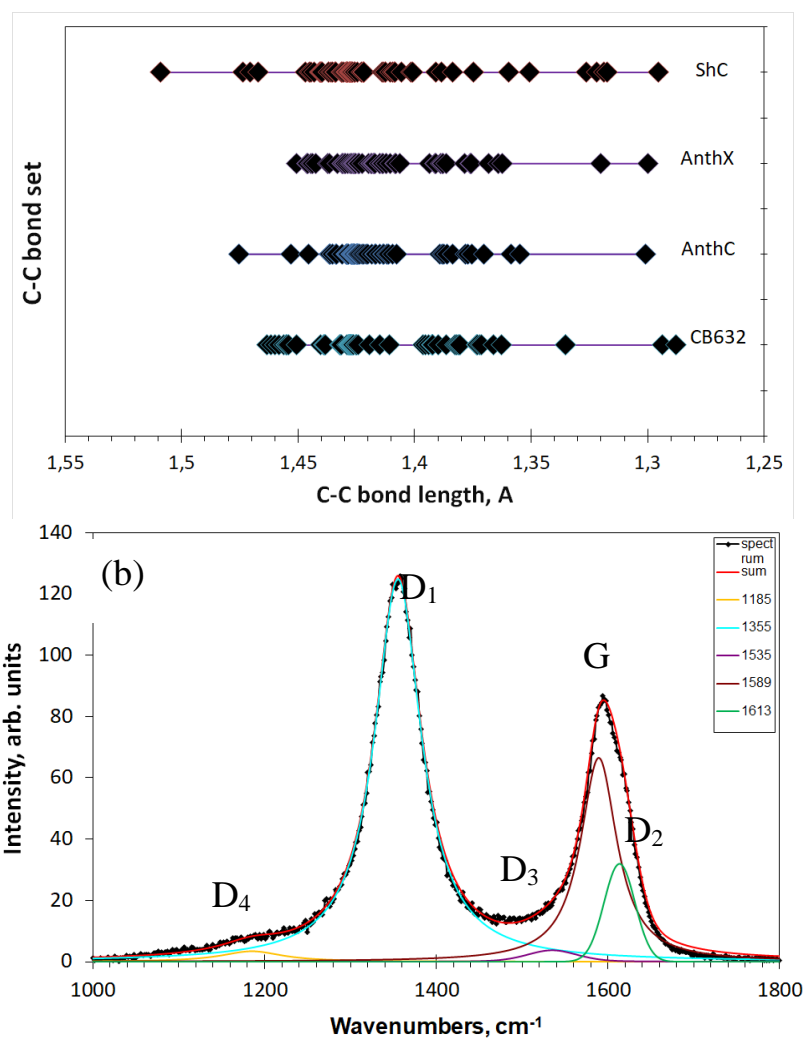

**Figure 8.** C=C bond sets of the model graphene domains (top) and deconvoluted Raman spectrum of shungite carbon (see text).

**Table 4**. Spectral characteristics of one-phonon Raman spectra of $sp^2$ amorphous carbons and a comparison of the data-treated and independently obtained size of graphene domains $L_a$

| Samples | FWHM G, cm$^{-1}$ | FWHM D, cm$^{-1}$ | $I_D/I_G$ [1] | $L_a$ according to Cançado et al. [77] | $L_a$ according to Tuinstra and Koenig [4] | $L_a$ XRD[2] |
|---|---|---|---|---|---|---|
| ShC [3] | 40–45 | 50–70 | 2.2–2.7 | 6.5–7 | 1.5–2 | 2.1 |
| AnthX [4] | 40–47 | 90–110 | 1.3–2.0 | 8–11 | 2–3.3 | 1.6 |
| AnthC | 40 | 100 | 2.0 | 8.3 | 2.1 | 2.1 |
| CB632 | 85 | 110 | 1.1 | 15 | 3.9 | 1.4 |
| CB624 | 34 | 45 | 0.6 | 29 | 7.6 | 14.6 |
| TE-rGO | 76 | 130 | 1.05 | 16 | 4.27 | >20 |
| AK-rGO | 56 | 93 | 2.2 | 7.5 | 2 | >20 |

[1] Integral intensities are considered.
[2] See sources of the data in Table 1.
[3] Data average over 7 samples [70].
[4] Data averaged over 8 samples [70].

As can be seen from the table, contrary to expectations from X-ray data, $L_a$, which should be the largest for shungites among the ScC-AnthX-AnthC-CB632 group, turns out to be the smallest obtained from Raman spectra, so the inverse X-ray dependence is observed. As seen from the table, the wished correspondence violates when the domain size is of the first tens of nanometers as well. In the present case, the situation does not seem unexpected, since, as was shown above, not only the size of graphene domain of the amorphics BSUs, but also the relevant heteroatom necklaces, as well as the BSUs packaging determine the RS intensity distribution between its constituent bands.

## VII. TWO-PHONON RAMAN SPECTRA OF $SP^2$ AMORPHOUS CARBONS

Second-order RSs of graphene materials have a long and rich history. Initially discovered in different graphite materials in the form of doublet of bands at ~ 2720 cm$^{-1}$ (strong) and 3248 cm$^{-1}$ (very weak) [78, 79], designated as G '(then 2D) and 2G, respectively, this region of the spectrum, demonstrating interesting properties, since that has been actively analyzed [80-83] (references are representative rather than exhaustive). Referring the reader to an excellent review of the state of art in this area [6], we will focus only on two features of the 2D band, which are important for further discussion. The first concerns the fact that there are no bands in the one-phonon spectrum of ideal graphite [79] and graphene [81] crystals, whose overtones or composites can be assigned as the 2D band. Only in graphite and/or graphene materials, which are obviously devoid of an ideal structure, the source of the overtone is attributed to the band D. This feature of "no one-phonon source" is tightly connected with the origin of the two-phonon spectrum of graphene materials. According to the fundamentals of Raman scattering [52], anharmonic dependence on normal coordinated of both vibrational and electronic energies (so called mechanic and electric anharmonicities) leads to the appearance of IR and Raman vibrational spectra beyond one-phonon. Mechanic anharmonicity is described with the third derivatives of potential energy while electric one results from the second derivatives of the object polarizability in Eq. (1). The two features influence the band-shape of the two-phonon spectra differently. Thus, the inconsistency of "no one-phonon source" and two-phonon spectra found in the case of RS of graphene crystals convincingly evidences a predominant role of the electric anharmonicity (*el*-anharmonicity below).

The second peculiarity of the 2D band concerns the surprising variability of its intensity with respect to the G band. So, in graphite crystal, the ratio of the total intensities $I_{2D}/I_G$ is ~ 1; in graphene crystal, it is equal to ~ 6 [82]; and in graphene whiskers it exceeds 13 [81]. Such large values and sharp fluctuations of the intensity, as well as the observation of 2D bands in graphite whiskers with an ideal crystal structure of not only intense 2D band, but also a broad high order RS located in the high-frequency region up to 7000 cm$^{-1}$, indicate an exceptionally big role for *el*-anharmonicity in this case. The authors are not aware of other examples of such a striking effect. Apparently, this property should be added to the treasure-box of graphene materials uniqueness.

Despite the huge number of publications concerning RS of graphene materials and the 2D band, in particular, the relationship of the 2D band with *el*-anharmonicity has not been considered. At the same time, modern vibrational spectroscopy attaches great importance to both mechanical and electrical anharmonicity, a joint participation of which provides good agreement between the experimental and calculated spectral data without fitting parameters [84–86]. Currently, this new approach can be applied to mid-size molecules such as thiophene or naphthalene, but the work on computational modules continues, so that graphene molecules of the first nm in size will be apparently able to be considered in the near future. Nevertheless, already obtained data on the quantitative accounting of anharmonicity in small molecules allow suggesting that the above-described behavior of the 2D band intensity in ideal crystals at a qualitative level is quite expected, if to assume the anharmonic behavior of the vibrational and electronic spectra of graphene-like structures to be peculiar. We dare to suggest that the highly delocalized character of the electron density of the graphene crystal and domains [87] contributes to such a pronounced anharmonicity. Moreover, the role of this feature in mechanical effect is evidently not direct. In contrast, the second derivatives of polarizability are directly determined by the state of the electronic system, which, possibly, determines the special role of *el*-anharmonicity in graphene. When the article was already written, it became known [88] that the intensity of G and 2D bands in graphene depends on the laser intensity in opposite way — it increases (G) and decreases (2D) with increasing the power, respectively. The authors explained the feature with asymmetric Fermi-Dirac distribution at the different optically resonant states contributing to Raman scattering stimulated with high electronic temperatures reached for pulsed laser excitation. Evidently, the distribution is tightly interrelated with the delocalized character of the electronic state.

Delocalization of electron density lays the foundation of specific properties of graphene molecules [89]. Supposing its particular role in the molecules RSs as well, it becomes obvious the observation of the second-order RS in such molecules as PHAs. It is this kind of spectrum that was recorded in the case of two PAH molecules [57] ($C_{78}$ and $C_{96}$ [62]). Looking at the spectra from this standpoint, we see that, actually, both spectra consist of the doublet of well-defined and narrow bands G and D located at 1603 and 1316 cm$^{-1}$, accompanied by an rather intense broad band with weakly expressed maxima at ~2610, ~2835, and ~2910 cm$^{-1}$. The first and last frequency markers are in good agreement with the frequencies of 2D and D+G bands. The appearance of the second maximum is obviously associated with the combination bands whose sources in the one-phonon spectrum have yet to be determined, thus supporting the *el*-anharmonicity origin of the spectra. In general, the band-shape of the two-phonon RSs of both molecules is similar to those of the studied $sp^2$ ASs shown in Fig. 9. Similarly to one-phonon RSs, the spectra are grouped in two columns related to natural (left) and synthetic (right) samples.

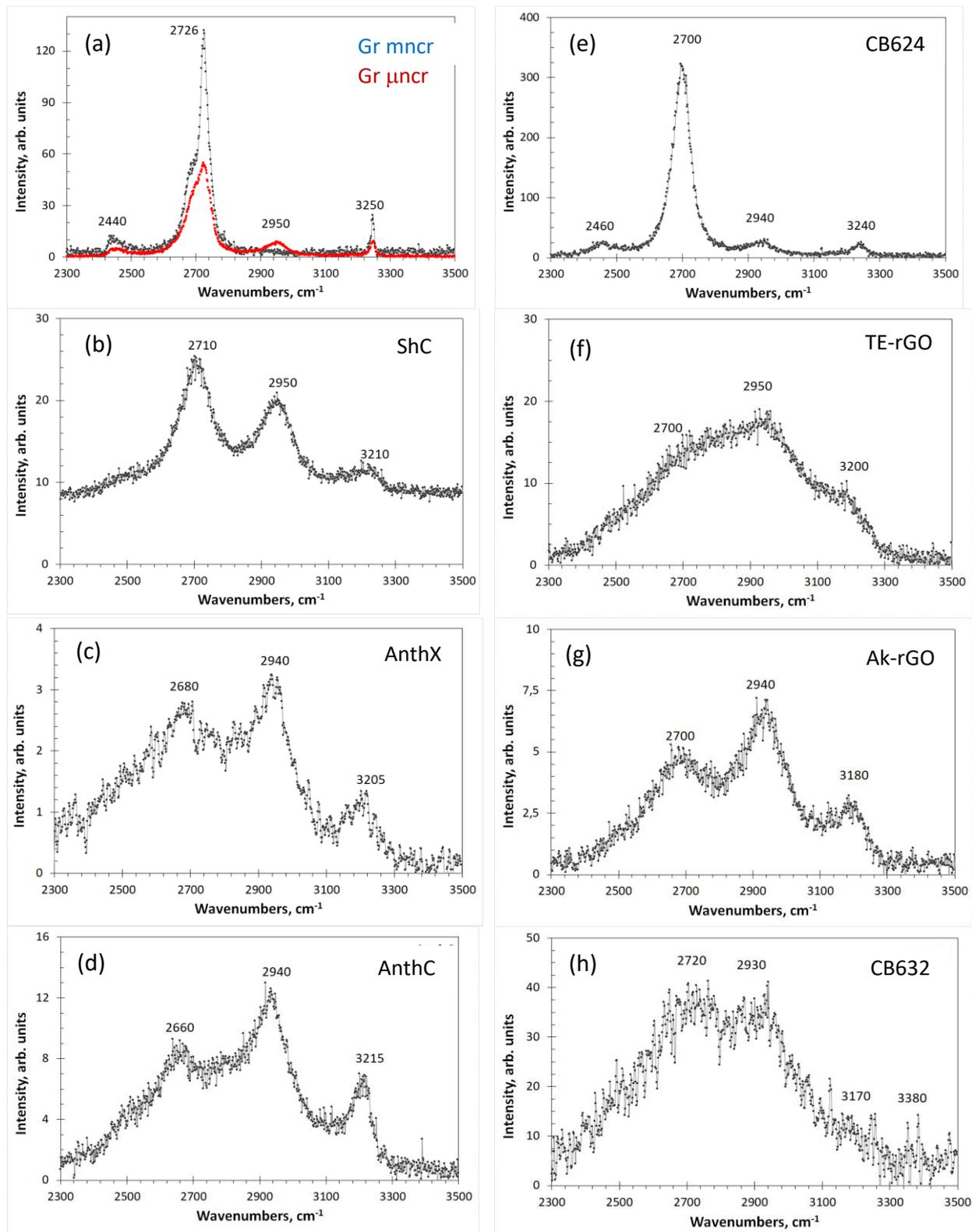

**Figure 9.** Two-phonon Raman spectra of *sp²* amorphous carbons. Sample marking is the same as in Fig. 5.

As seen in the figure, a characteristic four-component structure, clearly distinguished in the spectra of two graphites (Fig. 9a) and amorphous carbon black CB624 (Fig. 9e), can be traced in the spectra of all other amorphics. Three frequency markers at ~2700, ~2940, and ~3200 cm$^{-1}$, convincingly attributed to 2D, D+G, and 2G combinations, are steadily observed in all the spectra. The fourth marker at 2440 cm$^{-1}$, whose assignment remains unclear, is hidden in the low-frequency tails. Important to note that this marker involves at least one fundamental mode with frequency less than that of the lowest D mode. This mode has never been observed in the one-phonon spectra. This "no-one-phonon-source" feature is one more argument in favor of the *el*-anharmonic origin of the second-order Raman scattering of the studied *sp²* ACs.

Among the studied amorphics, the spectrum of black carbon CB624 occupies a central place. Thus, its close similarity with the spectrum of μncr graphite indicates that, despite the obvious difference in size of the short-range order region in the graphite and amorphic BSU, in both cases we are dealing with well-ordered structures making it possible to consider the spectra band-shape in the quasi-particle phonon approximation (widely reviewed in [82, 83]). As known from the solid-state physics, quasi-particle description, based on the conservation of translational symmetry, is applicable with respect to size-confined bodies when the size of the latter exceeds a critical value characteristic to the particle under consideration. In the case of phonons, the phonon free path determines this critical size (see the discussion of the issue for amorphics with molecular structure in [41, 90, 91]). The crystal-like band-shape of the RS of the carbon black CB624 under consideration evidences that the free path of optical phonons in graphene crystal is ~15 nm, as follows from Table 1. This conclusion is fully supported by the crystal-like behavior of one-phonon spectra of this solid shown in Fig. 5.

Since the theory of high-order RSs of large molecules is still being formed, a rigorous interpretation of their spectra, similar to proposed for one-phonon spectra in terms of molecular dynamics [5], is difficult. In connection with that, we propose to look at the spectra in Fig. 9 form the point of analytical triad: the size and heteroatom necklace of the relevant BSUs, as well as the BSUs package – as it was done with the one-phonon spectra in the previous section. As seen in the figure, the spectra form two distinguished groups, which join spectra of natural amorphics and Ak-rGO, on the one hand, and spectra of TE-rGO and CB632, on the other. This grouping replicates that one of one-phonon spectra in Fig. 5, clearly evidencing common reasons providing a significant likeness of the spectra within the group and a pronounced dislikeness between the groups. Obviously, the discussed features are in line with the concept on the governing role of the C=C BLD, thus allowing to attribute the RS shape of the first group solids to the effect of 'heteroatom necklace' of individual BSUs while that of the second group - to the BSUs turbostratic packing effect. Evidently, not the C=C BLD $\Delta\{l_{CC}\}$ itself, as in the case of one-phonon spectra, but its manifestation through the *el*-anharmonic action is responsible for the 2D spectra broadening. Despite the theory of *el*-anharmonic RS [84-86] has not so far allowed to formulate general regularities that govern the selection of particular modes for two-phonon spectra, it is evident that the action concerns a group of modes and when the modes pool is large, this leads to a significant broadening of the two-phonon bandshape. It should be noted that the two-phonon spectra turned out to be more sensitive to the triad components than one-phonon ones.

Concluding the analysis, one more thing should be mentioned. The frequency range of the studied two-phonon spectra coincides with the region of characteristic group frequencies (GFs) [92, 93], related to C-H stretching vibrations. Since, as seen in Table 2, all the studied amorphics, besides carbon blacks, are significantly hydrogenated, we tried to find the evidence of the hydrogen presence. However, none of the features related to the spectra observed can be attributed to characteristic GFs, such as 3050 cm$^{-1}$ related to methine groups [52] of natural amorphics as well as 2870-2970 cm$^{-1}$ and 2920-2980 cm$^{-1}$ of methylene and methyl groups [52] of technical graphenes TE-rGO and Ak-rGO, respectively. If hidden inside the broad bands, they might be revealed by a particular technique to be developed.

## VIII. CONCLUSION

Long-time and numerous studies of amorphous solids have led researchers to the conclusion that a confident interpretation of their Raman spectra is inseparable from the consideration of the nature and type of the solids amorphization. In the current work, these two aspects were considered together with respect to *sp²* ACs. Previous structural and analytical studies of these solids were analyzed from this viewpoint, due to which it was established that they can be attributed to molecular amorphics of a new type of amorphicity, which can be called *enforced fragmentation* of honeycomb canvas. Chemical reactions occurred at the fragment edges are suggested to be one of the most important causes of the fragmentation stabilization. Thus originated fragments, become the basic structural units (BSUs) of AC, are of particular kind presenting size-restricted graphene domains in the halo of heteroatom necklaces. The weak wdW interaction between the BUSs makes them the main defendants for the numerous properties of solids, including their IR absorption and Raman scattering.

Raman spectra of *sp²* ACs are considered at the molecular level, which is perfectly suitable to case. The molecular approximation is confirmed by a detail similarity of the spectra of the studied solids and 2D PAHs studied earlier. The similarity itself evidences a governing role of graphene domains, of both BSUs and PAHs, due to which a standard D-G-2D image of the spectra the graphene domain signature and remains until the structure is fully destroyed. The theory of Raman spectra of PAHs molecules, proposed by a team of Italian authors and confirmed in a number of computational experiments of the authors related to the PAHs experimental spectra, laid for the approach foundation. The nanometer and nano-submicron sizes of BSUs do not yet allow such experiments to be replicated in a worthy manner. Because of this, it was proposed to extend the conceptual results of the theory approved for PAHs to the case of the BUSs of the studied amorphics, supplementing analysis of their RSs by the concept of the C=C bond length dispersion (BLD), explaining the appearance of the D band as well as the spectra broadening as a whole. Besides, the available structure and chemical content data of the studied amorphics allowed suggesting an analytical triad: the size and heteroatom necklace of the BSUs, as well as the BSUs packing – as an additional tool for the *sp²* ACs RSs interpretation. This approach was applied to the interpretation of the one-phonon (D-G) and two-phonon (2D) spectra of the studied solids separately.

*One-phonon spectra*. Evaluated computationally, the C=C BLD for the studied amorphics constitutes ~0.2 Å attributing to the change of bond length from 1.5 Å to 1.3 Å. The corresponding dispersion of the C=C stretching frequencies is of ~400 cm$^{-1}$ from 1300 cm$^{-1}$ to 1700 cm$^{-1}$, thus covering the main spectral region of the amorphics RSs. The distribution of the bonds within the dispersion depends on the BSU size, the composition of heteroatom necklace and the BSUs multilayer structure. In light of this triad, the shape of the D-G spectra of natural amorphics and chemically reduced technical graphene is provided with the BSU size, over which a significant heteroatom-necklace effect is put on. In addition to this, the spectra of temperature-shock exfoliated technical carbon and one of the engineered carbon blacks exhibit a marked contribution of the BSU packing effect caused by turbostratic composition of the BSUs.

*Two-phonon spectra*. The C=C BLD concept allows explanation of the broadband 2D spectra of the studied solids as well, albeit not so straightforwardly as in the case of one-phonon ones since the spectra origin is provided with the electric anharmonicity of the BSU molecules. Nevertheless, size, heteroatom necklace, and packing effects are definitely exhibited in this case as well, thus making the total D-G-2D spectra of *sp²* ACs highly characteristic graphene domain signature with respect to the solids origin, production, and storage. It is this last circumstance that connects details of the Raman spectra of the AC BSUs in question with the history of the origin and production of the studied solids.

**Acknowledgements**

The authors are thankful to N.N. Rozhkova, S.V. Tkachev and V.M. Mel'nikov for supplying with samples of shungite carbon and technical graphenes as well as with necessary information concerning the sample production. Particular gratitude to S.I. Isaenko for assistance in Raman spectra recording and B.A. Makeev for XRD measurements. The study was performed as a part of research topics of the Institute of Geology of Komi Science Center of the Ural Branch of RAS (No. AAAA-A17-117121270036-7). The publication has been prepared with the support of the "RUDN University Program 5-100".

**Statement about conflicts**

There are no conflicts of interests between the authors